\documentclass[preprint,superscriptaddress,amsmath,amssymb,aps,prresearch]{revtex4-2}

\usepackage{graphicx} 
\usepackage{dcolumn}  
\usepackage{bm}       
\usepackage[hidelinks]{hyperref} 
\usepackage{placeins} 
\usepackage{tikz}

\begin{document}

\title{Scalable Quantum Error Mitigation with Physically Informed Graph Neural Networks}

\author{Huaxin Wang}
\affiliation{National Supercomputing Center in Zhengzhou, Zhengzhou University, Zhengzhou 450001, China}
\affiliation{School of Computer and Artificial Intelligence, Zhengzhou University, Zhengzhou 450001, China}

\author{Xinge Wu}
\affiliation{National Supercomputing Center in Zhengzhou, Zhengzhou University, Zhengzhou 450001, China}
\affiliation{School of Computer and Artificial Intelligence, Zhengzhou University, Zhengzhou 450001, China}

\author{Jiajun Liu}
\affiliation{National Supercomputing Center in Zhengzhou, Zhengzhou University, Zhengzhou 450001, China}
\affiliation{School of Computer and Artificial Intelligence, Zhengzhou University, Zhengzhou 450001, China}

\author{Ruiqing He}
\affiliation{School of Communication and Artificial Intelligence, School of Integrated Circuits, Nanjing Institute of Technology, Nanjing 211167, China}

\author{Jiandong Shang}
\affiliation{National Supercomputing Center in Zhengzhou, Zhengzhou University, Zhengzhou 450001, China}

\author{Hengliang Guo}
\affiliation{National Supercomputing Center in Zhengzhou, Zhengzhou University, Zhengzhou 450001, China}

\author{Qiang Chen}
\email{Corresponding author: qiangchen@zzu.edu.cn} 
\affiliation{National Supercomputing Center in Zhengzhou, Zhengzhou University, Zhengzhou 450001, China}

\begin{abstract}
Quantum error mitigation (QEM) provides a practical route for estimating reliable observables on noisy intermediate-scale quantum (NISQ) devices. Traditional QEM strategies, including zero-noise extrapolation (ZNE) and Clifford data regression (CDR), rely on noise scaling or global regression, and their performance is constrained by the exponential growth of the system degrees of freedom. We construct a graph-enhanced mitigation (GEM) framework, which incorporates physical information into the model representation. In this work, quantum circuits are encoded as attributed graphs. Hardware-level physical information is mapped to node and edge features: local noise parameters such as calibration parameters $T_1$, $T_2$, and readout errors are encoded at nodes, while coupling-related information such as two-qubit gate errors is encoded as edge features. Graph neural networks (GNNs) are used to model how errors propagate along the physical coupling structure and build up into non-local correlations. This allows the model to capture local interactions and part of the resulting non-local correlations across qubits. A dual-branch affine correction is applied to maintain consistency with physical constraints. Experiments on 10-qubit and 16-qubit random circuits executed on superconducting quantum processors show that GEM provides a level of accuracy comparable to CDR at small scales, while yielding lower mean absolute error and improved stability in zero-shot transfer to larger systems. Results of the traditional QEM strategy indicate that global regression methods remain effective in low-dimensional settings but become less reliable as system degrees of freedom grow. In contrast, GEM makes use of local physical structures to show better scalability and generalization, while preserving the overall error propagation patterns. This work provides a scalable and physically informed approach to error mitigation, enabling consistent performance across system sizes under limited classical resources.
\end{abstract}

\maketitle

\section{\label{sec:intro}Introduction}

Quantum computing has developed into a promising computational framework, with potential advantages in simulating quantum many-body systems, quantum chemistry, and quantum dynamics \cite{Kandala2017, Farhi2014,Choquette2021}. Current noisy intermediate-scale quantum (NISQ) devices \cite{Preskill2018}, however, are limited by physical noise, including energy relaxation, dephasing, gate imperfections, and readout drift \cite{Krantz2019, Clarke2008}.  These noise sources originate locally but can spread through circuit operations and lead to correlated effects across qubits. Quantum error correction (QEC) provides a rigorous path toward fault-tolerant computation, but its implementation requires substantial qubit overhead and high-fidelity operations. These requirements place strong constraints on both hardware scale and error rates \cite{Fowler2012, Google2023}, making large-scale deployment difficult in the NISQ regime. Under these conditions, Quantum error mitigation (QEM) has become an important approach for improving experimental accuracy without additional quantum resources \cite{Cai2023, Kandala2019}.

Several QEM methods have been developed, including zero-noise extrapolation (ZNE) \cite{Kandala2019, Li2017, GiurgicaTiron2020, Temme2017}, probabilistic error cancellation (PEC) \cite{Temme2017, VanDenBerg2023}, and Clifford data regression (CDR) \cite{Czarnik2021, Strikis2021, Lowe2021}, which perform well under specific noise assumptions. ZNE assumes that noise can be scaled in a controlled manner---a premise suitable for continuously adjustable noise strengths---but it remains susceptible to the amplification of shot noise in deep circuits. Conversely, PEC constructs unbiased estimators based on a detailed characterization of noise channels, but requires an exponential number of samples. Clifford data regression (CDR) utilizes supervised learning on classically simulable circuits, performing optimally for shallow or near-Clifford structures; however, as the system scales, the exponential expansion of the degrees of freedom imposes severe dimensionality constraints on its global mapping capabilities. Despite their theoretical foundations, these approaches encounter significant limitations in experimental settings. In realistic superconducting devices, noise often exhibits non-local correlations \cite{Maciejewski2021, Harper2023,Sung2021}, and many QEM methods involve exponential sampling costs \cite{Tsubouchi2023, Quek2024, Takagi2023}. Local interactions induce correlations between distant qubits. By encoding hardware connectivity and calibration parameters into graph representations and modeling them through message passing, GEM captures both local correlations and the resulting non-local effects.

Machine learning has recently been explored as a tool for studying quantum many-body systems and noise dynamics \cite{Carleo2017, Carrasquilla2017}, providing an alternative route for QEM. Neural error mitigation (NEM) was introduced by Bennewitz \textit{et al.} in 2022 to improve the estimation of ground states and observables in quantum simulations \cite{Bennewitz2022}. In 2024, Liao \textit{et al.} evaluated several machine learning models, including linear regression, random forests, multilayer perceptrons (MLPs), and graph neural networks (GNNs), and showed that machine learning-based QEM (ML-QEM) can significantly reduce mitigation cost \cite{Liao2024}. Tailored probabilistic machine learning frameworks utilizing Gaussian process regression have  been proposed to mitigate noise by capturing global parameter correlations within VQEs \cite{Jiang2024}. More recently, Patil \textit{et al.} investigated GNN-based approaches for QEM within variational quantum eigensolvers (VQE). Their work aims to achieve higher precision in molecular energy estimation through the development of graph neural models (GNM) \cite{Liao2024, Patil2025}. These works demonstrate the potential of learning-based approaches by exploiting circuit connectivity or ansatz structure to predict noiseless observables. By adopting graph representations, these methods effectively suppress expectation value errors while minimizing computational overhead. However, existing graph-based methods are typically tailored to specific ansatz structures and do not incorporate dynamic calibration parameters of the underlying hardware into the model representation. More importantly, these methods do not explicitly model how errors propagate along hardware connections and lead to correlated effects across the circuit. They are generally restricted to in-distribution generalization at a fixed system size and do not address the problem of scaling, namely how to transfer models trained on classically simulable subsystems to larger quantum circuits without retraining.

To achieve scalable and hardware-aware error mitigation, we propose the graph-enhanced mitigation (GEM) framework, which maps quantum circuits into interaction graphs with explicit physical content. In this representation, physical qubits and their couplings are mapped to nodes and edges, respectively, while local calibration parameters, such as $T_1$, $T_2$, and readout errors, together with two-qubit gate errors, are encoded as node and edge features. This construction embeds both local noise sources and their coupling structure into the model input. In this framework, the message-passing process of GNNs models the local propagation of noise along the physical coupling graph, while allowing the model to capture part of the resulting correlated effects across qubits. This mechanism provides a discrete approximation to error propagation in open quantum systems. Through local feature aggregation, the model captures the spatial structure of hardware noise. To preserve the physical consistency of corrected observables, we adopt a dual-branch affine transformation that combines structural embeddings with global statistical features.

We validate the proposed framework using 10-qubit and 16-qubit random circuits executed on the Tianyan-176 \cite{Tianyan2025} and Origin Wukong-180 superconducting quantum processors (QPU). Multiple baseline methods are included for comparison and ablation analysis. The results show that incorporating graph topology is necessary for capturing crosstalk effects, and that learning noise propagation from physically structured representations enables error mitigation that generalizes across system sizes without requiring exponential classical simulation.

This paper is organized as follows. Section II introduces the theoretical framework of GEM and the identity-preserving dual-branch formulation. Section III presents experimental results on real quantum hardware, including error mitigation across different physical levels, ablation studies, and zero-shot transfer across system sizes. Section IV discusses the implications of the proposed approach in terms of physical noise mechanisms and scalability.

\section{Theoretical Framework and Methodology}

Error processes in QPU arise from qubit-environment interactions and imperfect gate operations, and are described within the Lindblad master equation \cite{Pearle2012, Harper2023, VanDenBerg2023}. In practice, decoherence (e.g., $T_1$, $T_2$) and gate errors manifest predominantly as local processes that propagate through the physical coupling structure of the device \cite{Georgopoulos2021, Brand2024}. In multilayered random quantum circuits, errors diffuse through repeated entanglement operations, and their impact can be effectively characterized by the neighborhood structure.

Motivated by these considerations, we focus on two effective properties of noise in QPU: its predominantly local propagation constrained by device topology, and its slow temporal variation that can be captured by calibration parameters (including $T_1$, $T_2$, readout errors, and gate error rates). This leads to two verifiable hypotheses: first, incorporating coupling structure information should improve the prediction accuracy of local physical observables; and second, models utilizing calibration data should exhibit higher stability across operational fluctuations.

\begin{figure}[!htbp]
\centering
\includegraphics[trim={0cm 0cm 0cm 0.5cm},width=\columnwidth]{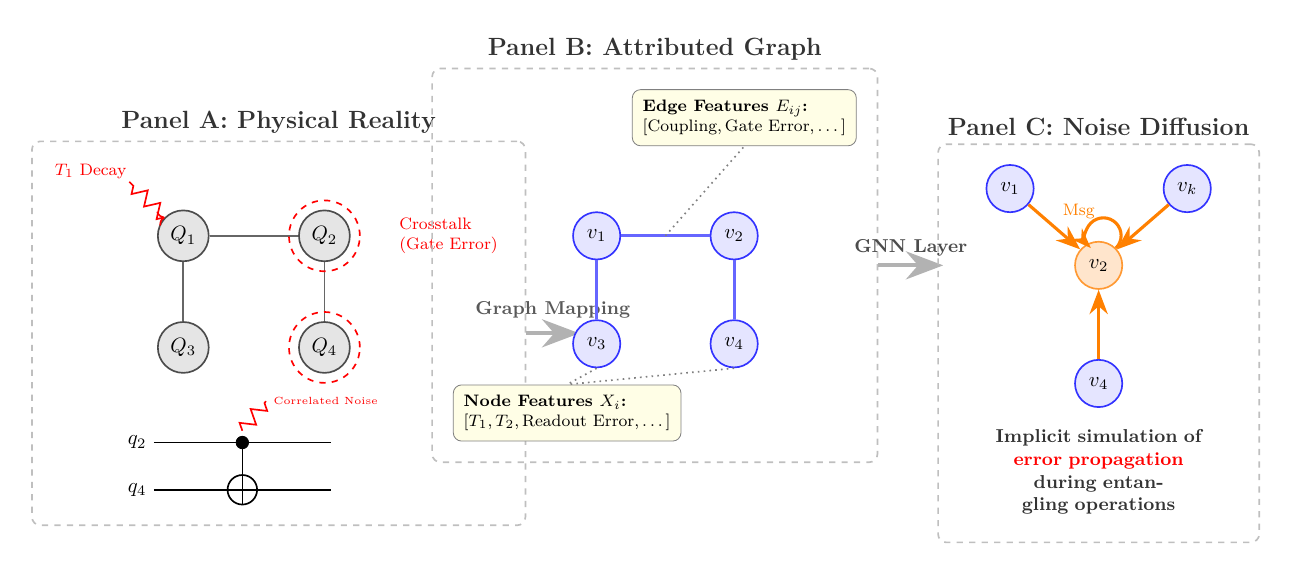} 
\caption{Schematic of the mapping from quantum circuits to physical attribute graphs. A shows the sources of physical noise in a QPU, where single-qubit decoherence ($T_1$/$T_2$) acts as a local effect, while two-qubit gate errors and crosstalk generate correlated noise along coupling edges. B provides the corresponding graph representation, encoding local and coupling-related noise features into nodes and edges. C demonstrates the GNN-based message-passing process used to characterize the local propagation of noise across the physical topology.}
\label{fig:1}
\end{figure}

Accordingly, a quantum circuit is represented here as a graph structure embedded with physical attributes (Fig.~\ref{fig:1}). Qubits and their coupling relationships are mapped to nodes and edges, respectively. Node features encode single-qubit calibration parameters, including $T_1$, $T_2$, and readout errors, while edge features characterize two-qubit gate errors. This representation explicitly embeds noise information within the topological structure, providing a physical inductive bias for subsequent learning. Furthermore, the neighborhood message-passing mechanism of GNNs, constrained by topology, aggregates node features layer by layer. This update process depends on local connectivity, formally corresponding to error propagation along coupling paths.

To accurately represent the mechanisms of hardware noise across different physical levels, this study refines traditional graph learning models by designing a dual-branch (global-local) physics-informed network architecture. This architecture addresses error characterization at two levels: local observables and global probability distributions. The former captures the local impact of noise on individual qubits, while the latter reflects the overall statistical structure of the many-body quantum state. Coupling these two types of information within a unified model allows for a more comprehensive description of spatial error propagation and its statistical accumulation.

\FloatBarrier

\subsection{Dataset construction}

When evaluating QEM methods on NISQ devices, the way training and test data are constructed plays a central role. It not only determines the generalization of the model, but also affects how faithfully the model reflects the underlying physical noise. If the dataset is restricted to specific circuit families, such as VQE, the model may overfit to particular gate patterns and lose applicability to more general circuits. To overcome this limitation, we generate random quantum circuits constrained by the native coupling topology of the hardware, ensuring both structural diversity and physical relevance.

\begin{figure}[!htbp]
\centering
\includegraphics[trim={0cm 19cm 0cm 3cm}, clip, width=\textwidth]{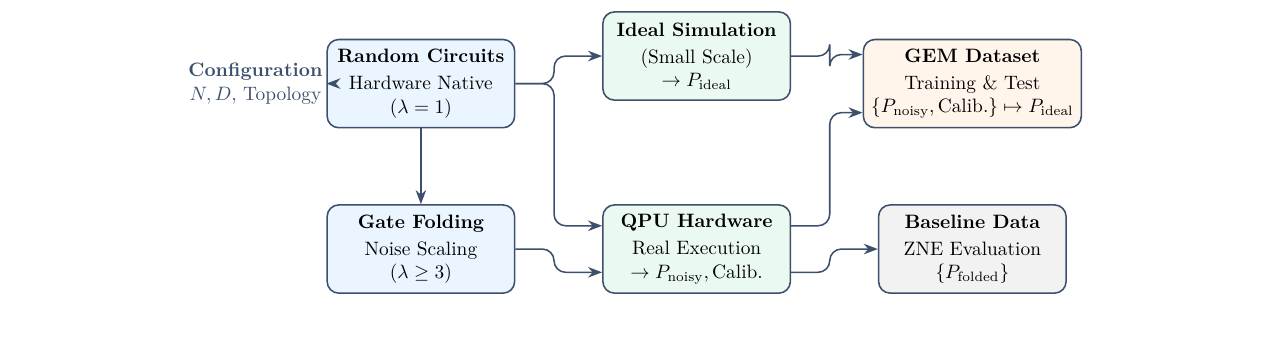} 
\caption{Dataset construction pipeline for hardware-aware random quantum circuits. Random circuits are executed on QPU to obtain the noisy empirical distribution $P_{\text{noisy}}$, while recording dynamic calibration parameters such as $T_1$, $T_2$, and readout errors. For 10-qubit systems, the ideal distribution $P_{\text{ideal}}$ is obtained via classical simulation and used as training labels. Additional data with multiple noise levels, generated through gate folding, are collected in parallel to support baseline methods such as ZNE.}
\label{fig:2}
\end{figure}

For a given number of qubits $N$ and circuit depth $D$, each circuit is constructed layer by layer. The overall pipeline is illustrated in Fig.~\ref{fig:2}. In each layer, single-qubit rotation gates are first applied to all qubits. The gate type is sampled uniformly from a set including $\{RX, RY, RZ, H, \dots\}$, with rotation angles drawn continuously from the interval $[0, 2\pi)$. Two-qubit gates are then applied by randomly selecting qubit pairs from the set of physically allowed couplings on the superconducting device, using gates such as CX or CZ. This procedure preserves randomness in entanglement generation while remaining consistent with hardware constraints, so that the resulting circuits can be executed directly without additional compilation, thereby avoiding extra compilation-induced noise.

\FloatBarrier

The dataset consists of 200 random quantum circuits for the 10-qubit system and 200 circuits for the 16-qubit system. For each circuit size, the data is split into a training set (80\%) and a test set (20\%). During data acquisition, each circuit is executed on real hardware with a fixed number of measurement shots to estimate the empirical output distribution $P_{\text{noisy}}$ in the computational basis. For classically tractable system sizes, such as 1-30 qubits, the same circuits are simulated without noise to obtain the reference distribution or target observables $P_{\text{ideal}}$. These ideal labels are used only during training. For larger systems, where classical simulation becomes intractable due to exponential cost, the model is evaluated directly without access to ideal labels, forming a strict zero-shot transfer setting.

To extend the coverage over different noise scales, additional circuits are generated using gate folding, for example with a threefold folding factor, to provide inputs for baseline methods such as ZNE. In contrast, the proposed GEM model uses only the original noisy circuits ($\lambda=1$) during both training and inference, without relying on artificially amplified noise. As a result, GEM estimates ideal observables from data at a single noise level, rather than requiring multiple noise-scaled measurements.

In addition to circuit structure and measurement outcomes, the dataset includes hardware calibration parameters recorded at each execution, such as the relaxation time $T_1$, coherence time $T_2$, and readout error rates. These quantities reflect the instantaneous coupling between qubits and their environment, and their temporal fluctuations constitute an important source of noise. By incorporating both circuit-level observations and hardware state information, the model can account for the nonstationary nature of noise during training. Table \ref{tab:1} illustrates the representative features and data structure utilized in this study.

\begin{table}[!htbp]
\centering
\caption{Sample dataset table for GEM covering both observable-level and distribution-level tasks.}
\label{tab:1}
\begin{ruledtabular}
\begin{tabular}{cccccc}
Circuit & Depth & Task & Input (noisy)  & Calibration& Graph Info \\ \hline
Circuit & Depth & Task & Input (noisy)  & Calibration & Graph Info \\ \hline
S01 & 5 & Distribution & $P_{\text{noisy}}(x)$  & ($T_1, T_2,\dots$) & Node features/Edge errors \\
S02 & 10 & Observable & $z_{\text{noisy}}$  & ($T_1, T_2,\dots$) & Node features/Edge errors \\

$\vdots$ & $\vdots$ & $\vdots$ & $\vdots$ & $\vdots$ & $\vdots$ \\

S199 & 50 & Distribution & $P_{\text{noisy}}(x)$  & ($T_1, T_2,\dots$) & Node features/Edge errors \\
S200 & 25 & Observable & $z_{\text{noisy}}$  & ($T_1, T_2,\dots$) & Node features/Edge errors \\
\end{tabular}
\end{ruledtabular}
\end{table}

Overall, the dataset combines three components: circuit topology, measurement distributions obtained from real hardware, and time-dependent calibration parameters. This integration of multiple information sources provides a basis for physically informed error modeling.

\FloatBarrier

\subsection{Physical information-supported graph modeling and GEM framework}

To formalize the quantum error mitigation problem, consider an ideal quantum circuit $C_{\text{ideal}}$ composed of $N$ qubits. In the absence of noise, its evolution corresponds to a unitary transformation, producing a probability distribution $P_{\text{ideal}}$ upon measurement. In practice, however, QPU provides access only to a noisy implementation, denoted as $C_{\text{noisy}}$, whose output distribution $P_{\text{noisy}}$ is affected by decoherence, gate imperfections, and readout errors. The central objective of error mitigation is therefore to recover physical observables from $P_{\text{noisy}}$ that approximate those of $P_{\text{ideal}}$, without introducing additional quantum resources.

\begin{figure*}[htbp]
\centering
\includegraphics[trim={0cm 0cm 0cm 0cm},clip, width=\textwidth]{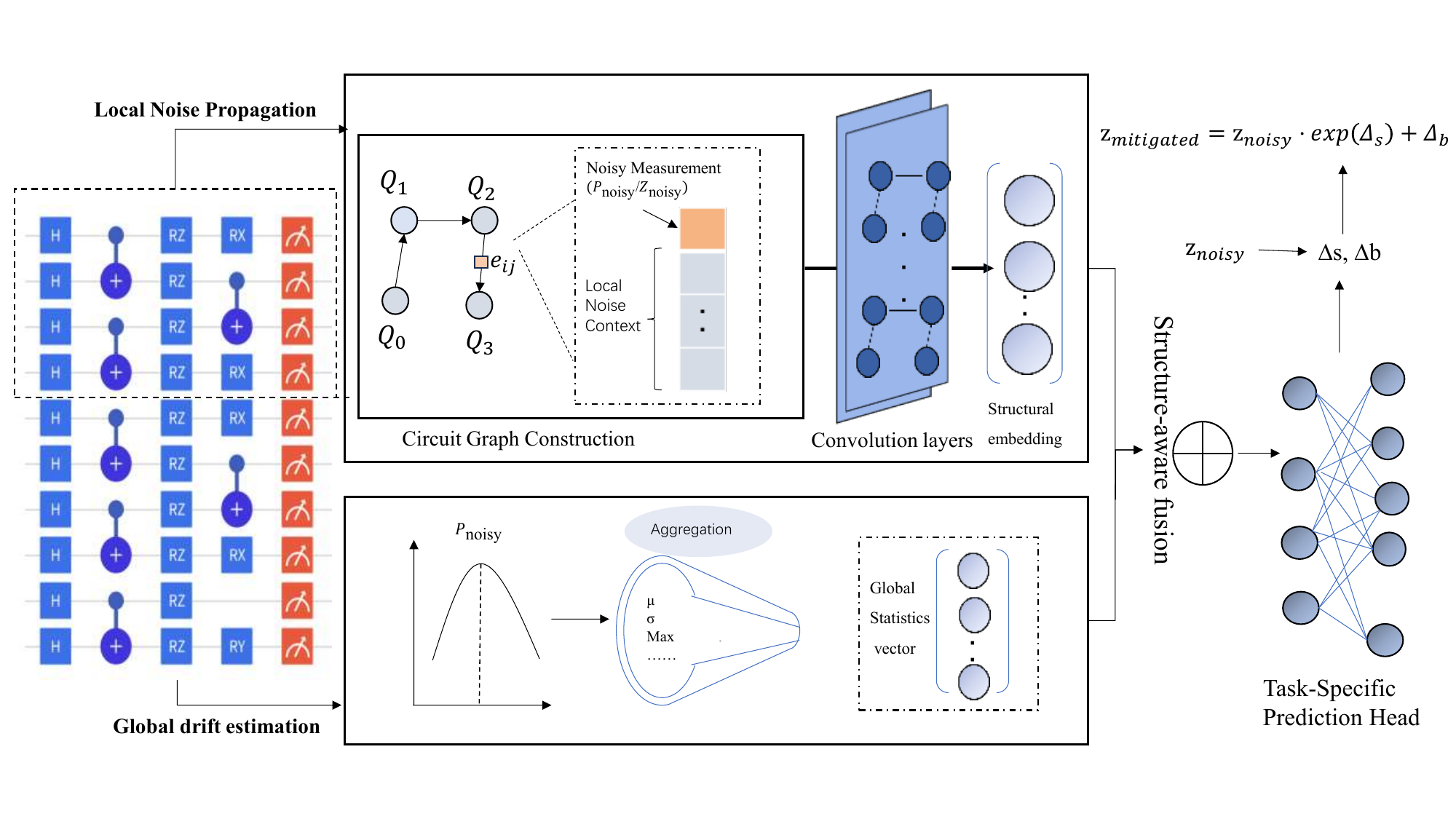} 
\caption{Overall architecture of the GEM framework. The model takes as input a graph representation of a quantum circuit defined on the physical coupling topology, $G=(V,E)$. Through iterative message passing over local neighborhoods, the graph neural network (GNN) captures the propagation of local noise. In parallel, global quantities derived from measurement statistics are incorporated to describe slowly varying system-level noise during experimental execution. After feature fusion, the model applies a correction to the raw noisy observables, yielding mitigated physical results. In this work, $\bm{z}$ represents the vector of quantum observables, which includes both local observables and global distribution statistics.}
\label{fig:3}
\end{figure*}

Based on this formulation, we introduce a physically informed graph-enhanced error mitigation framework, referred to as GEM. In this framework, quantum circuits are represented as attributed graphs with explicit hardware semantics, and graph neural networks (GNNs) are employed to model the propagation of noise over the underlying physical topology. The overall architecture of the model is illustrated in Fig.~\ref{fig:3}.

\FloatBarrier

In contrast to conventional approaches that rely on global noise scaling assumptions or direct regression between noisy and ideal observables, the proposed GEM framework is constructed from the perspective of noise processes in open quantum systems. Quantum circuits are represented as attributed graphs with explicit physical meaning, thereby introducing an inductive bias grounded in hardware structure. Specifically, a quantum circuit is mapped to an attributed directed graph
\begin{equation}
G = (V, E, X, E_f),
\end{equation}
where the node set $V$ corresponds to physical qubits and the edge set $E$ represents two-qubit operations and underlying couplings. The node feature matrix $X$ encodes local information, including measured expectation values and hardware calibration parameters such as $T_1$, $T_2$, and readout errors. Since relaxation and dephasing processes typically follow exponential decay, these parameters are transformed (for example, logarithmically or exponentially) prior to input, to reduce numerical instability caused by different dimensions.

Edge feature $E_f$ characterizes the error rate of two-qubit gates, thereby explicitly introducing nearest-neighbor crosstalk information. Apart from the local structure, to describe system-level noise drift, a global statistical vector $s$ is also defined, where $\phi (\cdot)$ represents a statistical aggregation operator on the observation data of the entire circuit (such as mean, variance, and extreme values). This vector is used to capture the macroscopic noise background caused by control electronics, temperature fluctuations, and so on.

Based on the above graph representation, GEM employs graph neural networks for feature extraction. Node representations are iteratively updated through a multi-layer message-passing mechanism, whose general form is
\begin{equation}
h_i^{(l+1)} = \sigma \left ( W h_i^{(l)} + \sum_{j \in \mathcal{N}(i)} \psi \big ( h_i^{(l)}, h_j^{(l)}, e_{ij} \big) \right),
\end{equation}
where $\mathcal{N}(i)$ denotes the neighborhood of node $i$, and $\psi (\cdot)$ is a nonlinear function conditioned on edge features. This structure constrains information flow along physical coupling edges, consistent with the locality of noise propagation in QPU. Since decoherence and crosstalk primarily propagate through neighboring interactions, this message-passing process can be viewed as a discrete approximation to a local noise propagation operator.

After multiple layers, a graph-level pooling operation produces a structural embedding of the circuit,
\begin{equation}
z_G = \mathrm{Pool}\left ( \left\{ h_i^{(L)} \right\} \right),
\end{equation}
which is concatenated with the global statistical vector $s$ to form a joint representation. This combined feature captures both local noise context and system-level drift, providing a unified input for prediction.

At the prediction stage, GEM does not directly regress ideal observables. Instead, it formulates error mitigation as a correction to noisy measurements. The model outputs parameters of an affine transformation applied to the input observable $z_{\text{noisy}}$,
\begin{equation}
z_{\text{mitigated}} = z_{\text{noisy}} \cdot \exp (\Delta_s) + \Delta_b,
\end{equation}
where $\Delta_s$ and $\Delta_b$ denote scale and bias corrections, respectively. The exponential form ensures that the scaling factor remains positive, preventing unphysical transformations of probabilities or expectation values.

A key component of this design is the identity mapping constraint. At initialization, output-layer parameters are set close to zero, such that $\Delta_s \approx 0$ and $\Delta_b \approx 0$, leading to $z_{\text{mitigated}} \approx z_{\text{noisy}}$. This ensures that training begins from the measured data and gradually learns noise-induced deviations. Compared with direct regression of ideal values, this formulation reduces the risk of numerical instability and keeps predictions within a physically meaningful range.

From a physical standpoint, this framework recasts error mitigation as learning the perturbative contribution of noise in an open quantum system. The GNN captures the spatial structure of noise propagation, while the affine transformation corresponds to a first-order correction to observables. This decomposition allows the model to learn effective noise behavior without assuming a specific parametric noise model.

Moreover, because message passing in GNNs depends only on local neighborhoods, the learned noise propagation patterns are independent of system size. This locality enables models trained on small-scale systems to be directly applied to larger circuits, maintaining effective error mitigation even in regimes where classical simulation is intractable.

\section{Random Circuit Error Mitigation}

The experimental data in this work were collected from real sampling on the Tianyan-176 and Origin Wukong-180 QPU. To quantify the statistical fluctuations over time, the error bars in the figures represent the standard error of the mean (SEM) from multiple independent real machine runs.

\subsection{Error Mitigation Behavior in Small-Scale Systems}

We first evaluate the performance of error mitigation methods on 10-qubit random quantum circuits, considering both the global probability distribution reconstruction and the prediction of local observables. These tasks correspond to the recovery of global statistical structures and local physical quantities, respectively, and their differences provide an important reference for subsequent analysis.

In the probability distribution case, we apply an overall correction to the noisy measurement results in logarithmic ratio form, predicting:
\begin{equation}
\log\left (\frac{P_{\text{ideal}}(x)}{P_{\text{noisy}}(x)}\right),
\end{equation}
and use this to reconstruct the mitigated distribution. The distribution similarity is quantified using the classical fidelity:
\begin{equation}
F_C (P_{\text{mitigated}}, P_{\text{ideal}}) = \left (\sum_x \sqrt{P_{\text{mitigated}}(x) P_{\text{ideal}}(x)}\right)^2,
\end{equation}
and the infidelity is used as the error metric:
\begin{equation}
\text{Infidelity} = 1 - F_C,
\end{equation}
where $F_C \in [0, 1]$ and $F_C = 1$ indicates perfect reconstruction of the ideal statistics. $\text{Infidelity} \in [0, 1]$, with $\text{Infidelity} = 0$ indicating perfect recovery. A lower infidelity value reflects better error mitigation performance.

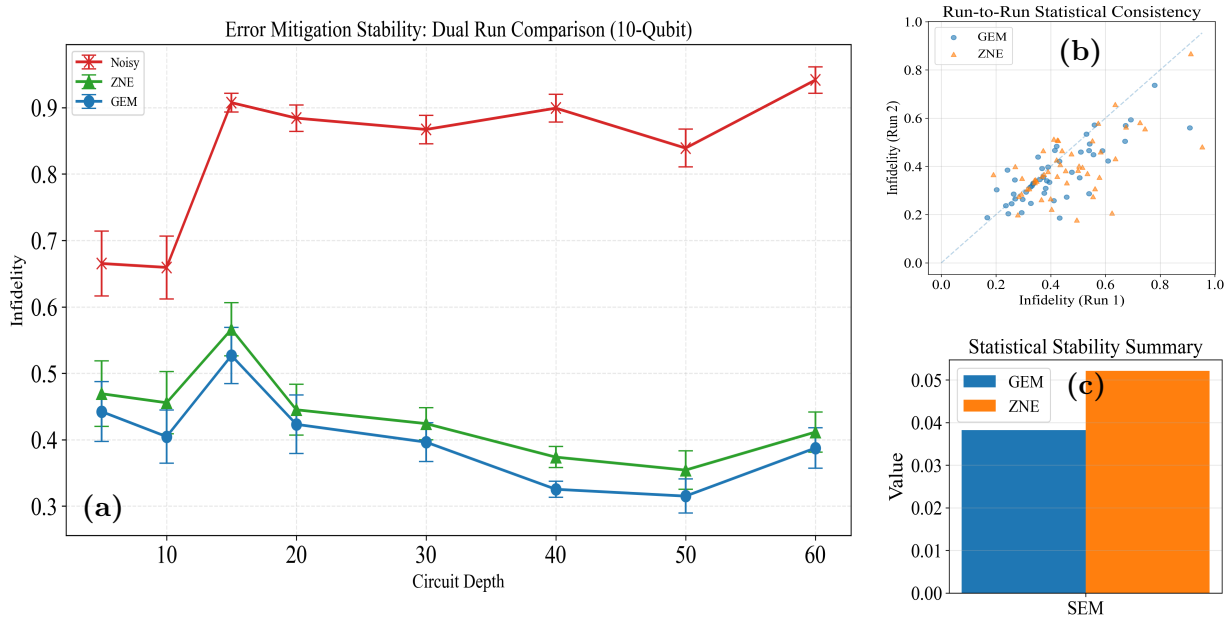
\begin{figure}[!htbp]
\centering

\begin{minipage}[t]{0.7\textwidth}
    \vspace{0pt} 
    \begin{tikzpicture}
        \node[inner sep=0pt] (figA) at (0,0) {\includegraphics[width=\textwidth, height=8.0cm]{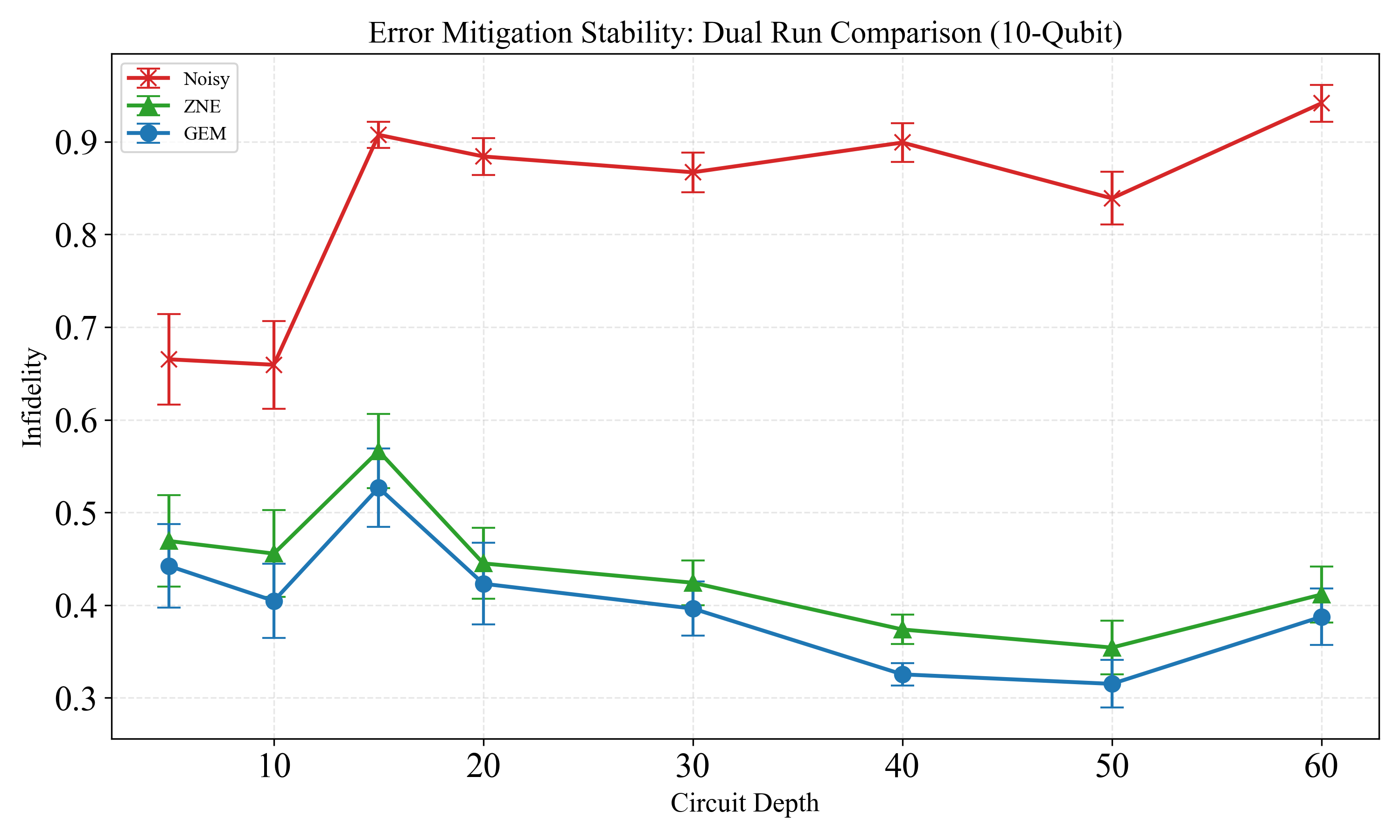}};
        
        \node[anchor=south west] at ([xshift=1cm, yshift=1cm]figA.south west) {\textbf{(a)}};
    \end{tikzpicture}
\end{minipage}
\hfill
\begin{minipage}[t]{0.29\textwidth}
    \vspace{0pt} 
    
    \begin{tikzpicture}
        \node[inner sep=0pt] (figB) at (0,0) {\includegraphics[width=\textwidth, height=4cm, trim=0 14 0 13,
    clip]{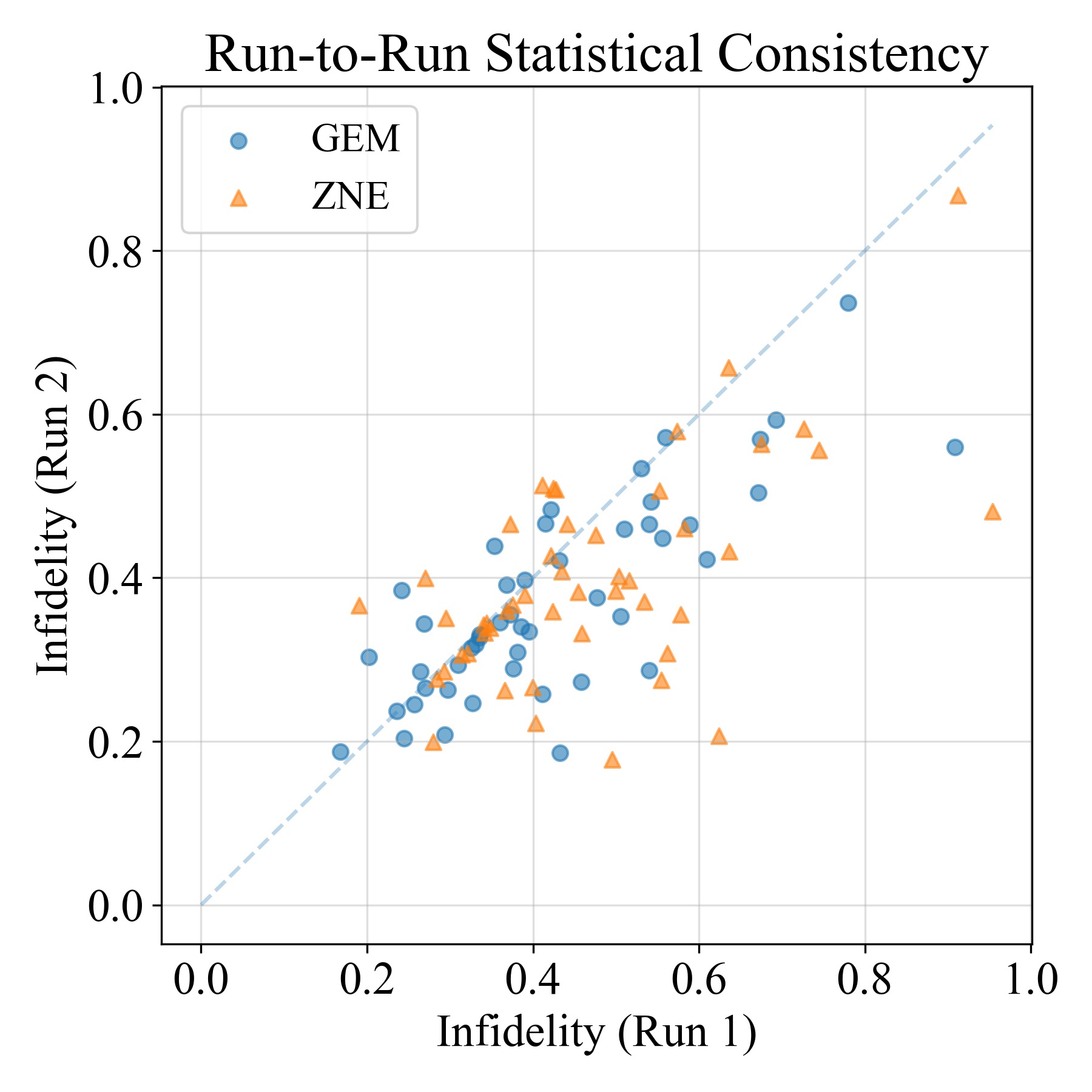}};
        
        \node[anchor=north west] at ([xshift=2.3cm, yshift=-0.3cm]figB.north west) {\textbf{(b)}};
    \end{tikzpicture}
    
    \vspace{0.4cm} 
    
    \begin{tikzpicture}
        \node[inner sep=0pt] (figC) at (0,0) {\includegraphics[width=\textwidth, height=3.7cm, trim=0 14 0 14,
    clip]{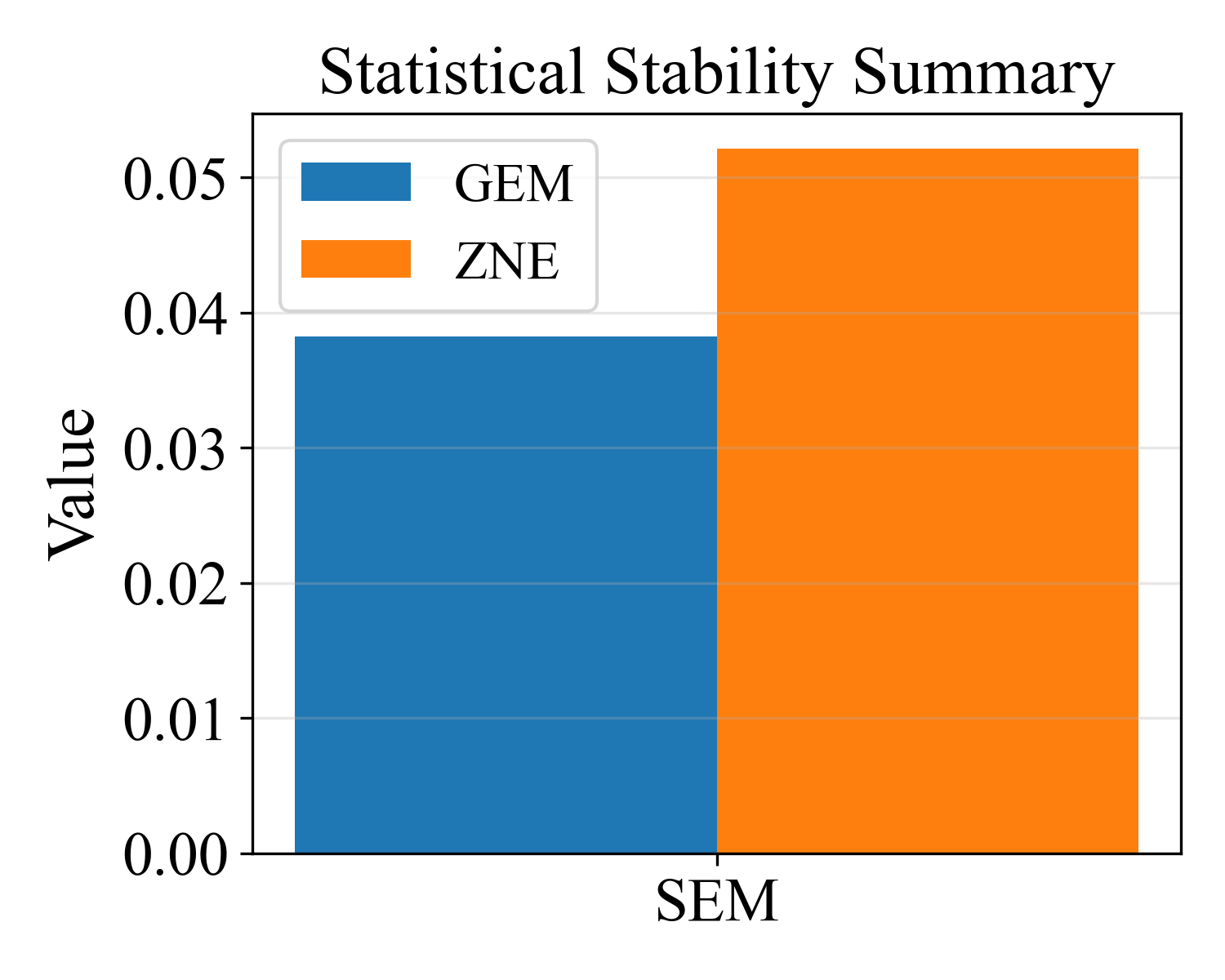}};
        
        \node[anchor=north west] at ([xshift=2.4cm, yshift=-0.3cm]figC.north west) {\textbf{(c)}};
    \end{tikzpicture}
\end{minipage}

\caption{Error mitigation performance and statistical stability on 10-qubit random circuits. (a) Depth-dependent infidelity. (b) Cross-run consistency of mitigation results. (c) Statistical fluctuations characterized by SEM.}
\label{fig:4}
\end{figure}


Figure \ref{fig:4}(a) shows that as circuit depth increases, the error of different methods accumulates in a consistent manner, with GEM showing a generally lower error than ZNE. To assess stability against temporal hardware drift, Figure \ref{fig:4}(b) compares independent real machine runs on the same set of circuits at different times. The average difference in infidelity between two runs is less than 17.5\%, with a Pearson correlation coefficient of $r = 0.797$ ($p < 0.05$). Figure \ref{fig:4}(c) illustrates cross-run fluctuations, where GEM's average error bar is notably lower (0.0382) than ZNE's (0.0521), indicating that GEM exhibits less dispersion between different runs.

In contrast to the global task of probability distribution, the local observable error mitigation more directly reflects the structure of local noise. We further investigate the performance of different methods on the same dataset using the Pauli-Z expectation value as a representative observable and use the mean absolute error (MAE) as the evaluation metric.
\begin{figure}[!htbp]
\centering
\includegraphics[width=0.67\textwidth]{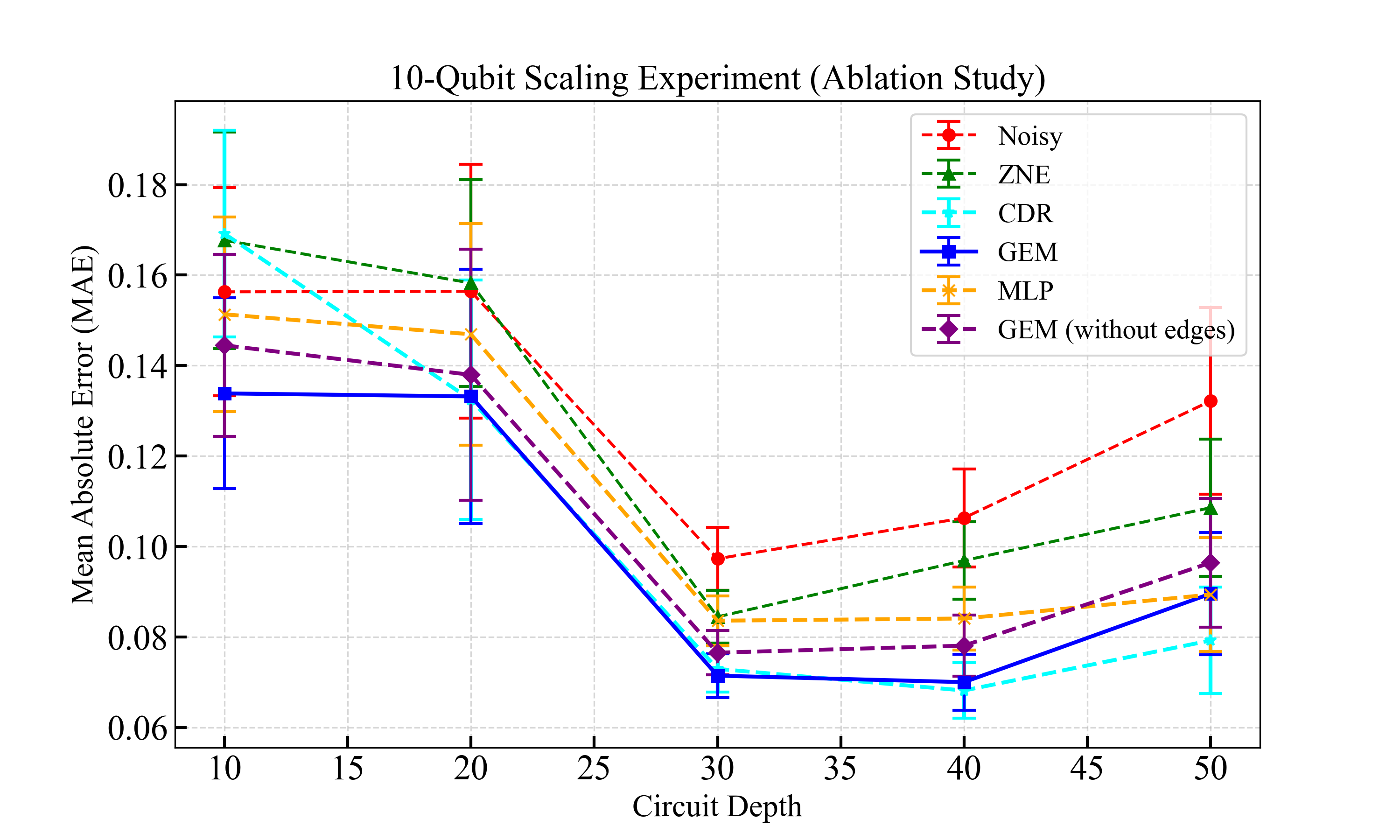} 
\caption{Error Mitigation Results for Observables on 10-Qubit Random Circuits}
\label{fig:5}
\end{figure}

\FloatBarrier

As shown in Figure \ref{fig:5}, in the lower depth regions of the circuit ($D \in [10, 20]$), the mean absolute error (MAE) of GEM is consistently lower than that of all comparison methods, including CDR and MLP. As the circuit depth increases ($D \gtrsim 40$), the relative performance of different methods changes: CDR begins to show progressively lower errors in the high-depth regions and outperforms GEM in certain intervals. This trend suggests that, in deep random circuits, the quantum state tends to become more mixed in a statistical sense, and local noise gradually spreads across multiple rounds of entangling operations, with its impact resembling a global effective mapping. In this case, CDR, based on global state regression, is more effective at fitting the overall relationship between noisy and ideal results.

\subsection{Cross-Scale Transfer and Scalability}

To systematically assess the differences in error mitigation behavior across quantum system scales, we design two complementary experiments: First, we retrain GEM on 16-qubit random circuits and compare it with the noisy baseline and ZNE; second, we apply models trained only on 10-qubit data directly to the 16-qubit system for zero-shot transfer testing.
\begin{figure}[!htbp]
\centering

\begin{minipage}[t]{0.495\textwidth}
    \vspace{0pt} 
    \begin{tikzpicture}
        \node[inner sep=0pt] (figA) at (0,0) {\includegraphics[width=\textwidth, trim=16 0 16 0,
    clip]{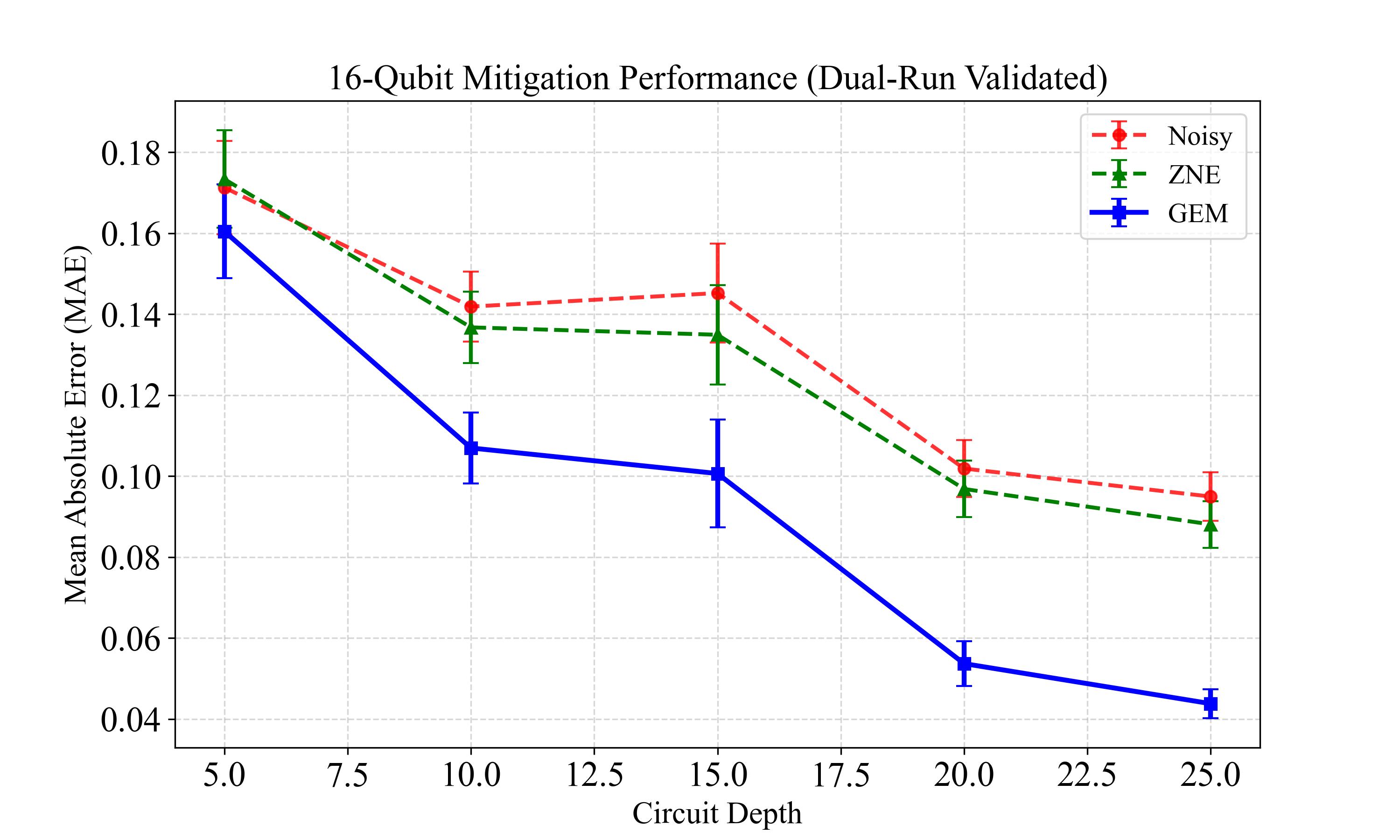}};
        
        \node[anchor=south west] at ([xshift=1.2cm, yshift=0.6cm]figA.south west) {\textbf{(a)}};
    \end{tikzpicture}
\end{minipage}
\hfill
\begin{minipage}[t]{0.495\textwidth}
    \vspace{0pt} 
    \begin{tikzpicture}
        \node[inner sep=0pt] (figB) at (0,0) {\includegraphics[width=\textwidth, trim=16 0 16 0,
    clip]{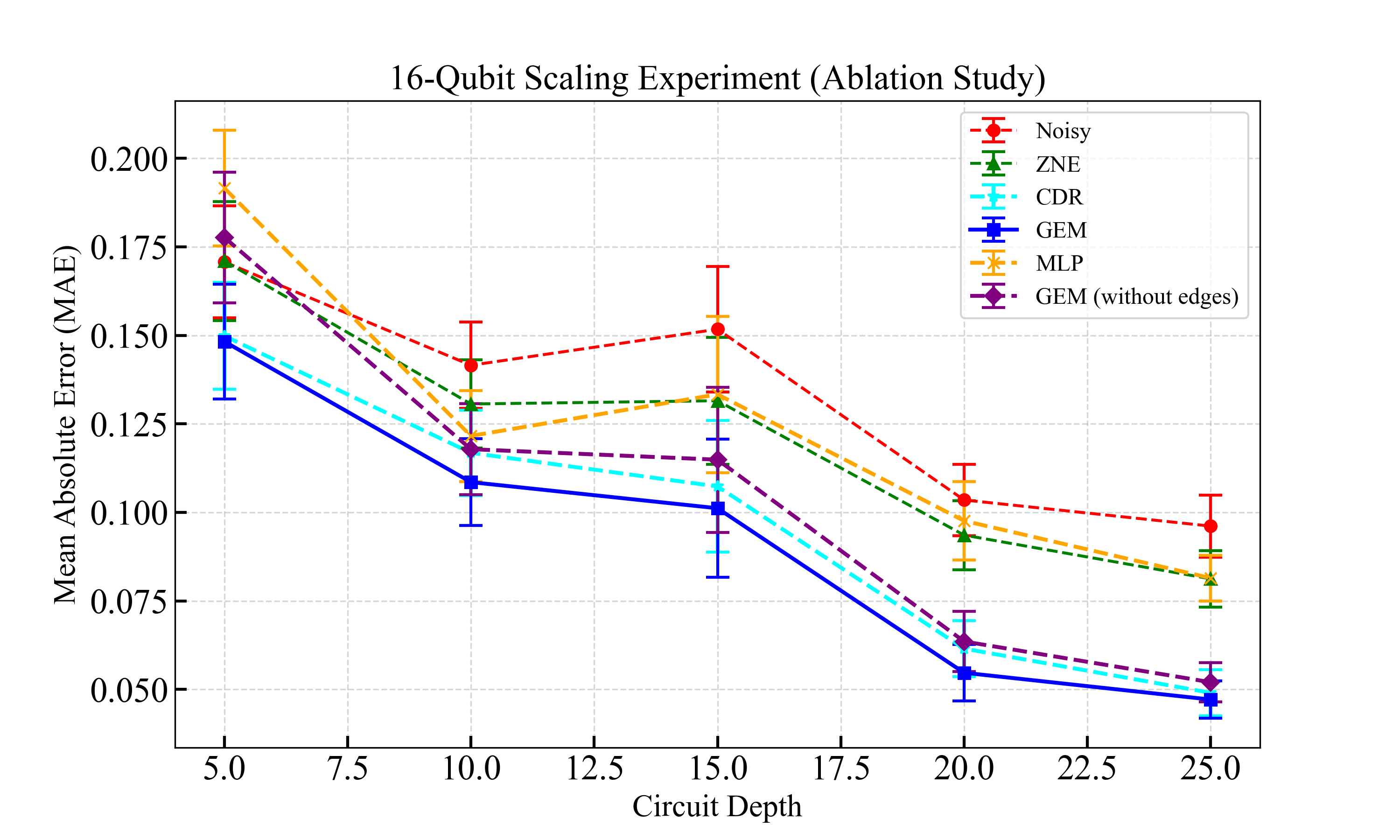}};
        
        \node[anchor=south west] at ([xshift=1.2cm, yshift=0.6cm]figB.south west) {\textbf{(b)}};
    \end{tikzpicture}
\end{minipage}

\caption{Cross-scale error mitigation performance. (a) MAE of GEM retrained on the 16-qubit system compared to Noisy and ZNE as a function of circuit depth. (b) Zero-shot transfer results of the 10-qubit trained model applied to the 16-qubit system.}
\label{fig:6}
\end{figure}

\FloatBarrier

\FloatBarrier

First, the results from retraining on the 16-qubit system are shown in Figure \ref{fig:6}(a). It can be observed that, for random quantum circuits, the MAE of all methods decreases with increasing circuit depth. This trend stems from the statistical nature of random quantum circuits: as circuit depth increases, the quantum state approaches a maximally mixed state \cite{Boixo2018}, causing the expectation values of local observables (such as $\langle Z \rangle$) to statistically converge to zero. 

The zero-shot transfer experiment in Figure \ref{fig:6}(b) further reveals how the methods generalize as the system size increases. When the model trained on 10-qubit data is directly applied to the 16-qubit system, all methods still follow the error evolution trend as the circuit depth increases, but the differences between models become more pronounced. GEM consistently maintains the lowest error across the entire depth range.

To compare the overall performance across various baselines, we perform global statistics on the error of the 16-qubit test sample, and the results are summarized in Table \ref{tab:2}.

\begin{table}[!htbp]
\centering
\caption{Overall Performance Statistics of Different Error Mitigation Methods on the 16-Qubit System}
\label{tab:2}
\begin{ruledtabular}
\begin{tabular}{lccc}
Method & Mean MAE & SEM & STD \\ \hline
Noisy & 0.1297 & 0.0059 & 0.1445 \\
ZNE & 0.1201 & 0.0060 & 0.1478 \\
MLP & 0.1243 & 0.0062 & 0.1524 \\
GEM (without edges) & 0.1036 & 0.0063 & 0.1542 \\
CDR & 0.0951 & 0.0055 & 0.1351 \\
GEM & 0.0903 & 0.0056 & 0.1391 \\
\end{tabular}
\end{ruledtabular}
\end{table}

\FloatBarrier

As shown in Table \ref{tab:2}, the overall performance differences between methods are clearer. Compared to the noisy baseline (Noisy), all error mitigation methods show some degree of error reduction. Among them, GEM achieves the best MAE (0.0903), slightly outperforming CDR (0.0951) and significantly outperforming ZNE (0.1201).

It is worth noting that the performance gap between GEM and CDR is relatively small in terms of numerical values, which is consistent with the observations made in small-scale systems. This reflects that, under the current data distribution, global regression methods can still approximate the ideal mapping to some extent. However, from the standard deviation (STD) and standard error of the mean (SEM), it is evident that both GEM and CDR perform similarly in terms of statistical stability, with GEM maintaining a lower baseline overall.

\section{Discussion}

The central premise of this work is to reinterpret noise in NISQ hardware not as a global black-box perturbation, but as a localized propagation process constrained by the physical coupling topology. Based on the dynamical picture of open quantum systems, we hypothesized that (i) errors predominantly propagate locally along the physical coupling topology, and (ii) noise exhibits slow temporal variation that can be effectively captured by calibration parameters. We examine these hypotheses in light of experimental observations and analyze how different mitigation strategies behave across physical levels and system scales.

\subsection{Error Structure Across Physical Levels}

Experimental results reveal distinct behaviors between global probability distributions and local observables, reflecting different physical manifestations of noise.

For probability distribution reconstruction in Fig.~\ref{fig:4}(a), all methods exhibit monotonically increasing error with circuit depth, indicating cumulative noise effects. Notably, GEM and ZNE follow nearly identical trends, differing mainly by a small, consistent offset (average improvement $<10.8\%$). This suggests that the overall trajectory of error evolution is governed by hardware noise itself rather than the mitigation method. In this sense, learning-based approaches do not alter the direction of error accumulation, but instead provide local corrections along an underlying physical trajectory.

This observation supports the locality hypothesis: error propagation is physically constrained by device topology. In NISQ devices, hardware noise generally consists of global correlated noise and local error mechanisms. Global noise presents as slow background fluctuations and lacks circuit-specific structure. Local errors, such as gate imperfections and crosstalk, originate strictly at specific interacting qubits (associated with two-qubit gates and coupling edges). Successive entangling operations cause these local errors to spread. Consequently, the final global error is not a sudden, system-wide event. It is simply the accumulation of local errors spreading through the physical connections. By mapping these local interactions, Graph Neural Networks trace the effective physical pathways of error spreading, thereby establishing a physically grounded representation that accounts for non-local correlations. The consistent trend between GEM and ZNE reflects this topological constraint.

Subsequent experiments on local observables and cross-scale transfer further validate this framework. Specifically, the depth-dependent crossover in local observable mitigation in Fig.~\ref{fig:5}---where GEM outperforms baselines in shallow circuits but is overtaken by CDR in deep ones—indicates error propagation is initially confined to local connections, before scrambling drives the system toward a maximally mixed state. Furthermore, zero-shot transfer in Fig.~\ref{fig:6} confirms GEM's consistent performance across system sizes. This demonstrates that the model decouples noise characterization from the global state, capturing scale-invariant local dynamics.

Hypothesis (ii), concerning slow temporal variation of noise, is supported by cross-run consistency analysis. As shown in Fig.~\ref{fig:4}(b), independent executions at different times yield strongly correlated results (Pearson $r = 0.797, p < 0.05$), indicating that the underlying noise structure remains stable over time. Furthermore, Fig.~\ref{fig:4}(c) shows that GEM reduces statistical fluctuations compared to ZNE (SEM: 0.0382 vs 0.0521). This stability stems directly from the explicit embedding of physical hardware information. By encoding dynamic calibration data—specifically relaxation ($T_1$) and coherence ($T_2$) times, along with readout error rates—as node features, and strictly enforcing logic-to-physical qubit mapping to align the circuit with the actual hardware topology, GEM captures time-dependent variations in noise. This physically grounded alignment effectively accounts for slow drift in hardware conditions.

In contrast, local observable estimation (Pauli-Z, Fig.~\ref{fig:5}) exhibits qualitatively different behavior. GEM consistently outperforms both MLP and its edge-ablated variant across all depths. Recent studies have demonstrated that errors in near-term quantum processors are rarely independent; mechanisms such as hardware crosstalk induce significant spatially correlated noise \cite{Harper2023, Maciejewski2021}. Despite recent advancements in robust error suppression, effectively mitigating these non-local correlated errors remains an acknowledged open challenge \cite{REAS2024}. While GEM does not completely resolve this complex issue, it provides a distinct structural advantage. Instead of treating non-local correlations as an unstructured global perturbation, GEM naturally accommodates their topological footprint by embedding physical interaction pathways directly into its graph topology. The necessity of this structural prior is evident in the ablation results: removing edge features leads to a clear degradation. This indicates that coupling structure is essential for capturing realistic noise, and further suggests that GEM captures the effective dynamics of non-local errors by tracing physical connectivity rather than assuming independent qubit noise.

CDR achieves comparable performance and slightly surpasses GEM in some deep circuits. This can be attributed to the increasing mixing of quantum states at larger depths: repeated entangling operations effectively spread local noise, making its effect resemble a global mapping. In a 10-qubit system, such mappings remain tractable, allowing global regression methods to approximate noise effects effectively. This explains why global methods can remain competitive in low-dimensional settings.

\subsection{Circuit Scalability of the GEM Framework}

As system size increases from 10 to 16 qubits, differences between methods evolve from quantitative to qualitative.

First, all methods exhibit decreasing MAE with circuit depth in 16-qubit systems in Fig.~\ref{fig:6}(a). This is not due to improved mitigation, but arises from the statistical nature of random circuits: deeper circuits approach near-maximally mixed states \cite{Boixo2018}, causing local observables (such as $\langle Z \rangle$) to concentrate near zero and reducing the absolute error scale. Within this regime, retrained GEM consistently achieves lower error than Noisy and ZNE across all depths, indicating that it removes systematic bias without altering the underlying statistical convergence. This suggests that the model performs constrained corrections rather than introducing new dynamics.

More pronounced differences emerge in zero-shot transfer in Fig.~\ref{fig:6}(b). When trained on 10-qubit data and applied directly to 16-qubit systems, all models preserve the same depth-dependent trend, but diverge significantly in error magnitude. GEM maintains the lowest error, while MLP and edge-ablated variants degrade noticeably. Table~\ref{tab:2} quantifies this effect: GEM achieves $\text{MAE} = 0.0903$, outperforming CDR (0.0951) and ZNE (0.1201), while removing edge features increases error to 0.1036. The consistency of this gap across depths indicates a systematic limitation rather than statistical noise.

This behavior can be understood through inductive bias. CDR relies on global regression from noisy to ideal states, which requires sufficient coverage of the underlying state space. As system size increases, the Hilbert space grows exponentially ($2^N$), making such mappings increasingly difficult to constrain, and thus limiting generalization \cite{McClean2018}. In contrast, GEM operates on local neighborhood interactions defined by the coupling graph. It models non-local correlations by following physical error propagation. Its predictions depend on local error propagation patterns rather than the full state space, making them inherently less sensitive to system size. The degradation observed when removing edge features further highlights that topology-encoded structure is essential for cross-scale generalization.

These results indicate a scale-dependent transition: global regression methods can approximate noise effectively in low-dimensional systems, but lose effectiveness as dimensionality increases, where explicit structural modeling becomes necessary. This is consistent with the underlying physical assumption that noise propagation is governed by local coupling rather than global state structure.

\section{Conclusion}

In this work, we proposed a physically informed graph-based error mitigation framework (GEM) and validated it experimentally on a QPU. By encoding quantum circuits as attributed graphs and combining a local--global dual-branch architecture, the method provides a unified representation of both local noise and cross-qubit correlated errors. GEM framework embeds physical information into the model. A dual-branch architecture is used to maintain global consistency.  Experimental results show that GEM consistently outperforms unstructured learning models and zero-noise extrapolation (ZNE) in terms of overall error and statistical stability, maintaining robustness even in the 16-qubit zero-shot transfer setting. Notably, GEM significantly outperforms global regression methods  in shallow and medium-depth circuits. In these regimes, error propagation is primarily confined to local hardware connections. While global regression may be competitive in deep circuits under low-dimensional settings, its generalization degrades as system size increases due to the exponential growth of the Hilbert space. In contrast, graph-based approaches maintain consistent error propagation tracking and offer improved scalability by incorporating hardware-compliant locality.

While the current verification is limited to medium-scale random circuits ($\leq 16$ qubits) and single-body Pauli-$Z$ observables, future work can proceed along two directions. First, the framework can be extended to many-body observables by treating qubit pairs as local substructures and introducing subgraph-level aggregation to model cross-qubit correlations. Higher-order correlation functions can be implemented by extending the message-passing range, testing the model's ability to express non-local errors without altering its overall structure. Second, extending this framework to non-random circuits (such as variational quantum circuits) is essential. In such scenarios, errors are tightly coupled with the circuit structure and entanglement patterns, which significantly affect optimization performance, such as gradient decay~\cite{Wang2021, Robbiati2026}. By comparing mitigation performance on random circuits with structured circuits, future work can directly evaluate the generalization ability of the graph model under diverse error distributions.

\section*{Acknowledgments}

This work is supported by the National Key Research and Development Program of China (2024YFB4504104). The quantum computing tasks in this work were implemented on superconducting quantum processors provided by China Telecom and Origin Quantum. This work is also supported by Jiangsu Province Engineering Research Center of IntelliSense Technology and System.


\begin{thebibliography}{99}


\bibitem{Kandala2017}
A. Kandala \textit{et al.}, Hardware-efficient variational quantum eigensolver for small molecules and quantum magnets, Nature \textbf{549}, 242 (2017).

\bibitem{Farhi2014}
E. Farhi, J. Goldstone, and S. Gutmann, A quantum approximate optimization algorithm, arXiv preprint arXiv:1411.4028 (2014).

\bibitem{Preskill2018}
J. Preskill, Quantum computing in the NISQ era and beyond, Quantum \textbf{2}, 79 (2018).

\bibitem{Krantz2019}
P. Krantz \textit{et al.}, A quantum engineer's guide to superconducting qubits, Appl. Phys. Rev. \textbf{6}, 021318 (2019).

\bibitem{Clarke2008}
J. Clarke and F. K. Wilhelm, Superconducting quantum bits, Nature \textbf{453}, 1031 (2008).

\bibitem{Fowler2012}
A. G. Fowler \textit{et al.}, Surface codes: Towards practical large-scale quantum computation, Phys. Rev. A \textbf{86}, 032324 (2012).

\bibitem{Google2023}
Google Quantum AI \textit{et al.}, Suppressing quantum errors by scaling a surface code logical qubit, Nature \textbf{614}, 676 (2023).

\bibitem{Cai2023}
Z. Cai \textit{et al.}, Quantum error mitigation, Rev. Mod. Phys. \textbf{95}, 045005 (2023).

\bibitem{Kandala2019}
A. Kandala \textit{et al.}, Error mitigation extends the computational reach of a noisy quantum processor, Nature \textbf{567}, 491 (2019).

\bibitem{Li2017}
Y. Li and S. C. Benjamin, Efficient variational quantum simulator incorporating active error minimization, Phys. Rev. X \textbf{7}, 021050 (2017).

\bibitem{GiurgicaTiron2020}
T. Giurgica-Tiron \textit{et al.}, Digital zero noise extrapolation for quantum error mitigation, in \textit{2020 IEEE International Conference on Quantum Computing and Engineering (QCE)} (IEEE, 2020).

\bibitem{Temme2017}
K. Temme, S. Bravyi, and J. M. Gambetta, Error mitigation for short-depth quantum circuits, Phys. Rev. Lett. \textbf{119}, 180509 (2017).

\bibitem{VanDenBerg2023}
E. van den Berg \textit{et al.}, Probabilistic error cancellation with sparse Pauli–Lindblad models on noisy quantum processors, Nat. Phys. \textbf{19}, 1116 (2023).

\bibitem{Czarnik2021}
P. Czarnik \textit{et al.}, Error mitigation with Clifford quantum-circuit data, Quantum \textbf{5}, 592 (2021).

\bibitem{Strikis2021}
A. Strikis \textit{et al.}, Learning-based quantum error mitigation, PRX Quantum \textbf{2}, 040330 (2021).

\bibitem{Lowe2021}
A. Lowe \textit{et al.}, Unified approach to data-driven quantum error mitigation, Phys. Rev. Res. \textbf{3}, 033098 (2021).

\bibitem{Maciejewski2021}
F. B. Maciejewski \textit{et al.}, Modeling and mitigation of cross-talk effects in readout noise with applications to the Quantum Approximate Optimization Algorithm, Quantum \textbf{5}, 464 (2021).

\bibitem{Harper2023}
R. Harper and S. T. Flammia, Learning correlated noise in a 39-qubit quantum processor, PRX Quantum \textbf{4}, 040311 (2023).

\bibitem{Tsubouchi2023}
K. Tsubouchi, T. Sagawa, and N. Yoshioka, Universal cost bound of quantum error mitigation based on quantum estimation theory, Phys. Rev. Lett. \textbf{131}, 210601 (2023).

\bibitem{Quek2024}
Y. Quek \textit{et al.}, Exponentially tighter bounds on limitations of quantum error mitigation, Nat. Phys. \textbf{20}, 1648 (2024).

\bibitem{Takagi2023}
R. Takagi, H. Tajima, and M. Gu, Universal sampling lower bounds for quantum error mitigation, Phys. Rev. Lett. \textbf{131}, 210602 (2023).

\bibitem{Carleo2017}
G. Carleo and M. Troyer, Solving the quantum many-body problem with artificial neural networks, Science \textbf{355}, 602 (2017).

\bibitem{Carrasquilla2017}
J. Carrasquilla and R. G. Melko, Machine learning phases of matter, Nat. Phys. \textbf{13}, 431 (2017).

\bibitem{Bennewitz2022}
E. R. Bennewitz \textit{et al.}, Neural error mitigation of near-term quantum simulations, Nat. Mach. Intell. \textbf{4}, 618 (2022).

\bibitem{Liao2024}
H. Liao \textit{et al.}, Machine learning for practical quantum error mitigation, Nat. Mach. Intell. \textbf{6}, 1478 (2024).

\bibitem{Patil2025}
S. Patil, D. Mondal, and R. Maitra, Machine learning approach toward quantum error mitigation for accurate molecular energetics, J. Chem. Phys. \textbf{163}, 024101 (2025).

\bibitem{Tianyan2025}
T. Q. Group, Tianyan: Cloud services with quantum advantage, arXiv preprint arXiv:2512.10504 (2025).

\bibitem{Pearle2012}
P. Pearle, Simple derivation of the Lindblad equation, Eur. J. Phys. \textbf{33}, 805 (2012).

\bibitem{Georgopoulos2021}
K. Georgopoulos, C. Emary, and P. Zuliani, Modeling and simulating the noisy behavior of near-term quantum computers, Phys. Rev. A \textbf{104}, 062432 (2021).

\bibitem{Brand2024}
D. Brand, I. Sinayskiy, and F. Petruccione, Markovian noise modelling and parameter extraction framework for quantum devices, Sci. Rep. \textbf{14}, 4769 (2024).

\bibitem{Klimov2018}
P. V. Klimov \textit{et al.}, Fluctuations of energy-relaxation times in superconducting qubits, Phys. Rev. Lett. \textbf{121}, 090502 (2018).

\bibitem{Boixo2018}
S. Boixo \textit{et al.}, Characterizing quantum supremacy in near-term devices, Nat. Phys. \textbf{14}, 595 (2018).

\bibitem{McClean2018}
J. R. McClean \textit{et al.}, Barren plateaus in quantum neural network training landscapes, Nat. Commun. \textbf{9}, 4812 (2018).

\bibitem{Wang2021}
S. Wang \textit{et al.}, Noise-induced barren plateaus in variational quantum algorithms, Nat. Commun. \textbf{12}, 6961 (2021).

\bibitem{Odake2025}
T. Odake \textit{et al.}, Robust error accumulation suppression for quantum circuits, Phys. Rev. Res. \textbf{7}, 033029 (2025).

\bibitem{Jiang2024}
T. Jiang, J. Rogers, M. S. Frank, O. Christiansen, Y.-X. Yao, and N. Lanatà, Error mitigation in variational quantum eigensolvers using tailored probabilistic machine learning, Phys. Rev. Res. \textbf{6}, 033069 (2024).

\bibitem{Choquette2021}
A. Choquette \textit{et al.}, Quantum-optimal-control-inspired ansatz for variational quantum algorithms, Phys. Rev. Res. \textbf{3}, 023092 (2021).

\bibitem{Robbiati2026}
M. Robbiati \textit{et al.}, Real-time error mitigation for variational optimization on quantum hardware, Phys. Rev. Res. \textbf{8}, 013262 (2026).

\bibitem{REAS2024}
T. Odake \textit{et al.}, Robust error accumulation suppression for quantum circuits, Phys. Rev. Res. \textbf{7}, 033029 (2025).

\bibitem{Sung2021}
Y. Sung \textit{et al.}, Realization of high-fidelity CZ and ZZ-free iSWAP gates with a tunable coupler, Phys. Rev. X \textbf{11}, 021058 (2021).
\end{thebibliography}
\end{document}